\documentclass[showpacs,preprint,amssymb]{revtex4}
\usepackage{graphicx}
\usepackage{dcolumn}
\usepackage{bm}
\def\be{\begin{equation}} 
\def\ee{\end{equation}}
\def\bea{\begin{eqnarray}} 
\def\eea{\end{eqnarray}}
\def\line{\hbox to \hsize}    
\def\frac #1#2{{#1\over #2}}

\def\tr{{\rm  tr\,}}
\def\det{{\rm det\,}}

\def\det{{\rm det\,}}

\def \x{{\bf x}}

\def \p{{\bf p}}  
\def \a{{\bf a}}
\def \d{{\bf d}}

\def \ket #1{{\vert #1\rangle}}
\def \bra #1{{\langle #1\vert}}

\def\eval #1#2#3{{\langle#1\vert#2\vert#3\rangle}} 

\def\1{\mbox{\bf 1}}
\def\bm#1{\mbox{\boldmath$#1$}} 

\begin{document}

\title{Classical chiral kinetic theory and anomalies  in  even space-time dimensions}

\author{VATSAL DWIVEDI}

\affiliation{University of Illinois, Department of Physics\\ 1110 W. Green St.\\
Urbana, IL 61801 USA\\E-mail: vdwived2@illinois.edu}

\author{ MICHAEL STONE}

\affiliation{University of Illinois, Department of Physics\\ 1110 W. Green St.\\
Urbana, IL 61801 USA\\E-mail: m-stone5@illinois.edu}

\begin{abstract}  

We propose  a classical action for the motion  of massless Weyl fermions   in a background gauge field  in $(2N+1)+1$  spacetime dimensions. 
We  use this action to derive    the collisionless   Boltzmann equation for a  gas of such particles,  and show how   classical versions of the gauge and 
Abelian chiral anomalies arise from the   Chern character  of  the  non-Abelian Berry connection  that parallel transports the   spin degree of freedom in momentum space.  

\end{abstract}

\pacs{11.10.Kk, 11.15.-q, 12.38.Aw, 12.38.Mh, 71.10.Ca}

\maketitle

\section{Introduction}

The axial anomaly---a non-conservation of currents associated with chiral fermions---is usually  regarded as a purely quantum effect,   so  it is rather   surprising that it is possible  to extract   the abelian anomaly  from  a purely  classical Hamiltonian  phase-space    calculation \cite{stephanov}.   The authors of \cite{stephanov}  did so by considering the classical dynamics of  a finite-density gas of Weyl fermions in background electromagnetic  field. They observed   that  the  incompressibility of phase space   allows  the anomalous inflow    of  particles from the negative-energy Dirac sea into  the positive-energy Fermi sea \cite{haldane,son1,chen-pu-wang-wang}  to  be  reliably counted by keeping track of the density  flux   only  near the Fermi surface. Near the Fermi surface, and therefore   well   away from the dangerously quantum  Dirac point, a  classical  Boltzmann equation becomes sufficiently accurate  for this purpose. The only quantum input  required is knowledge of how to normalize the phase space measure  and  a simple   computation of the   momentum-space gauge field  that  accounts for the gyroscopic effect of the    Weyl particle's spin. This  gauge field is a Berry-phase  effect that  subtly alters the classical canonical structure so that ${\bf x}$ and ${\bf p}
$ are no longer conjugate variables,  and $d^3pd^3x$ is no longer the element of phase space volume \cite{niu,duval}.

In a previous paper \cite{stone-dwivedi} we extended the analysis  of  \cite{stephanov} and considered the motion of Weyl particles in a background {non-Abelian\/} gauge field.  In this way we obtained  the non-Abelian {gauge\/} anomaly in 3+1 dimensions. Our  calculation, like that in \cite{stephanov},   relied on a number of simplifications  peculiar to three spatial dimensions---in particular that the Berry phase   is indeed  a phase and not a more general unitary  matrix. In the present paper we derive both the Abelian and the gauge anomaly in any even number $(2N+1)+1 $ of space-time dimensions.  Apart from the  intrinsic interest of the role of higher-dimension anomalies in the hydrodynamics of a Weyl gas \cite{loganayagam} and other fluids with anomalies \cite{nair}, it turns out that the structure of the calculation becomes more transparent when we discard  the special features of  the lower dimensional dynamics.

In section \ref{SEC:action} we review the action that describes the motion of a Weyl particle in a background $3+1$ Abelian gauge field. We then explain how this action can be extended so as  obtain the  motion of a Weyl particle in any even-dimensional  spacetime, and  in a background  non-Abelian field.  In section \ref{SEC:liouville} we review  the differential form formulation of Hamiltonian dynamics and show how it is  modified  when the underlying symplectic form becomes time dependent.  We use this   language  to extend Liouville's theorem on phase-space volume conservation to the time dependent case, and identify the   vulnerabliity in the proof where permitting a singularity   allows for anomalous conservation laws.
In \ref{SEC:boltzmann} we temporarily assume that the time dependent Liouville theorem is {\it not\/} compromised and show how   it implies  the non-anomalous  conservation of the Abelian number current and the covariant conservation of the gauge current.   Section \ref{SEC:anomalies} then reveals  how the  non-trivial   Chern character of the  momentum space spin  connection  provides a source term in  Liouville conservation law. We then show that  this source term gives rise to both  the singlet anomaly and the covariant form of the gauge anomaly.  An appendix displays  the  efficiency of  the formalism  used in  the present paper   compared to that of  our previous  method. Two further   appendices provide technical details of results used in the main text.

 \section{A  classical action for  Weyl fermions}
 \label{SEC:action} 
 
 It was shown in \cite{stephanov} that 
 the classical phase-space action for a 3+1 dimensional Weyl fermion moving in   a  background electromagnetic  field  is 
 \be
S[{\bf x},{\bf p}] = \int  dt\left({\bf p}\cdot {\dot {\bf x}}-|{\bf p}| - e\phi({\bf x})+e{\bf A}\cdot \dot{\bf x}   -\a\cdot {\dot {\bf p}}\right).
\label{EQ:abelian-action}
\ee
  Here $e{\bf A}\cdot \dot {\bf x}$ is the standard  coupling of the Maxwell  vector potential ${\bf A}$ to the  velocity $\dot{\bf x}$ of the charge $e$ particle. The combination 
  $e\phi({\bf x},t)+|{\bf p}|$, where
 $\phi$ $(=-A_0)$  is the scalar potential and  $|{\bf p}|$ is the kinetic energy of the massless fermion,  is the classical Hamiltonian $H({\bf x},{\bf p})$.  The   $\a\cdot {\dot {\bf p}}$ term     
accounts for   the  gyroscopic effect of the spin angular momentum,  which for a  right handed  massless particle is forced  to point in the direction of the momentum.

\subsection{Berry connection in momentum space}
\label{SUBSEC:berry}

The  momentum-space gauge  field ${\bf a}({\bf p})$ appearing in (\ref{EQ:abelian-action})  is  the adiabatic Berry connection \cite{berry1} which has components  
 \be
 a_k=i \eval{{\bf p},+}{\frac{\partial}{\partial p^k}}{{\bf p},+}.
 \ee
These components  are  obtained from the $E=+|{\bf p}|$ eigenvector $\ket{{\bf p},+}$  of the  quantum  Hamiltonian
 \be
\hat H_{\bf p}={\bm \sigma}\cdot {\bf p}, 
\label{EQ:3dquantumH}
\ee 
and the  resulting  Berry   curvature   ${\bf b}= \nabla\times {\bf a} $ possesses  a monopole singularity 
\be
\nabla\cdot {\bf b}= 2\pi \delta^3({\bf p})
\label{EQ:monopole}
\ee
at ${\bf p=0}$.  
When  ${\bf p}\gg 0$ and  the $2|{\bf p}|$ energy gap is large  we can ignore the negative energy eigenstate  $ \ket{{\bf p},-}$ and safely make the adiabatic approximation that allows us to forget  most   quantum effects. We  retain only the  Abelian Berry phase, now interpreted   as a classical gauge potential. 
 The adiabatic approximation fails   near ${\bf p=0}$, however,  and (\ref{EQ:monopole}) should  be understood  only as  shorthand   indicating the presence of a    non-zero Berry flux through   surfaces surrounding (but distant from)  the origin \cite{stephanov}.

We desire  to generalize  (\ref{EQ:abelian-action})  from $3$ to $2N+1$ space  dimensions.  The principal complication is that in dimensions greater than three the  Berry {\it phase\/}    is replaced by a unitary evolution {\it matrix} and ${\bf a}({\bf p})$ by  a  non-Abelian Berry connection.  Extracting the effects of the Berry  connection   by taking a suitable  limit of the quantum system is complicated, and is ultimately equivalent to the conventional Feynman diagram calculations of  anomalies that we wish to avoid. Instead we build on the results of \cite{stephanov} by letting     symmetry and gauge invariance dictate the necessary  modification to  the Abelian action functional (\ref{EQ:abelian-action}).

In  $2N+1$ space dimensions the quantum Weyl Hamiltonian (\ref{EQ:3dquantumH}) becomes 
\be 
\hat H_{\bf p}= \sum_{i=1}^{2N+1} \Gamma_i p^i,
\ee
where the $\Gamma_i$ are a set of $2^N$-by-$2^N$ Dirac gamma matrices (or more accurately Dirac alpha matrices) obeying
\be
\{\Gamma_i,\Gamma_j\}= 2\delta_{ij}.
\label{EQ:clifford}
\ee
The gamma matrices will  also obey 
\be
 \Gamma_{1}\Gamma_{2}\cdots \Gamma_{{2N+1}} = \pm i^N {\mathbb I}_{2^N},
\ee
where  $\pm$ sign depends the Weyl particle's helicity---{\it i.e.\/}\  which of the two inequivalent  irreducible representations of the Clifford algebra (\ref{EQ:clifford}) is selected. 

The eigenvalues  of $\hat H_{\bf p}$ remain $\pm |{\bf p}|$, but   each energy level  is  now $2^{N-1}$-fold  degenerate. The  positive energy eigenspaces  $V_+({\bf p})$ form  the fibres of   a non-trivial ${\rm Spin}(2N)$ bundle over momentum space minus its  origin.    If $\ket{{{\bf p},\alpha,+}}$, $\alpha= 1,\ldots, 2^{2N-1}$ form  a basis for $V_+({\bf p})$,  the  natural non-Abelian
Berry connection  connection on the bundle  has components \cite{wilczek-zee}.
\be
{\mathfrak a}_{\alpha\beta,k} =  i  \eval{{\bf p},\alpha, +}{\frac{\partial}{\partial p^k}}{{\bf p},\beta,+}.
\ee
Its matrix-valued curvature tensor is 
\be
{ \mathfrak F}_{ij}= \frac{\partial {\mathfrak a}_j }{\partial p^i}- \frac{\partial{\mathfrak  a}_i}{\partial p^j} -i[{\mathfrak a}_i,{\mathfrak a}_j].
 \ee
We  anticipate that  ${\bf a}({\bf p})$  is to be replaced by the matrix-valued ${\mathfrak a}(\p)$.

It is often convenient to expand the matrix-valued connection and curvature as  
\bea 
{\mathfrak a}_i &=&\sum_{n<m}X_{nm} {\mathfrak a}^{nm}_i, \nonumber\\
{ \mathfrak F}_{ij}= &=& \sum_{n<m}X_{nm} { \mathfrak F}^{nm}_{ij},
\eea
where  $X_{nm}=-X_{mn}$   are  the  $2^{N-1}$-by-$2^{N-1}$ Hermitian matrix  generators  of    the   Lie algebra of ${\rm Spin}(2N)$.  The generator $X_{nm}$ corresponds  an infinitesimal rotation  in the $n,m$ plane, and the   set of all $2N^2-N$  pairs $(n,m) $, $n<m$, should be though of as the range of a  single index labelling the generators. We will  abbreviate this  index as $(n)$. For example the commutation relation 
\be
[X_{ij},X_{mn}]= i(\delta_{mi}X_{jn}- \delta_{mj}X_{in}   -X_{mi}  \delta_{jn}+ X_{mj}\delta_{in}),
\ee
will be abbreviated  as  
\be
[X_{(a)}, X_{(b)}] =i {f_{(a)(b)}}^{(c)}X_{(c)}.
\ee
In other words, ${f_{(a)(b)}}^{(c)}$ (with  indices in parentheses) denotes  the structure constants of the Lie algebra of ${\rm Spin}(2N)$.

In \cite{stone-dwivedi} we also  replaced the scalar-valued abelian   $(A_0, {\bf A})$ fields  in (\ref{EQ:abelian-action}) by  a non-Abelian Hermitian matrix  gauge field 
\be
A_\mu =\hat \lambda_aA^a_\mu, \quad 
F_{\mu\nu} = \frac{\partial A_\nu}{\partial x^\nu}- \frac{\partial A_\mu}{\partial x^\mu}-i[A_\mu,A_\nu]
\ee
for a compact simple gauge group $G$.
Here the $\hat \lambda_a$ are  the matrices  corresponding to  the generator $\lambda_a\in  {\rm Lie}(G)$ in the representation that determines the ``charge'' of our Weyl particle.  The generators obey 
\be
[\lambda_a,\lambda_b] =i{f_{ab}}^c \lambda_c.
\label{EQ:liealgebra}
\ee
Thus, ${f_{ab}}^c$ (indices without parentheses)  are the structure constants of ${\rm Lie}(G)$.
To obtain a classical version of the gauge and spin quantum degrees of freedom,  we   must somehow  replace the matrix-valued gauge and spin connection fields with scalar-quantities. This is  because  any  action functional must be the integral of scalar.  However the  simplest scheme of 
taking traces over the matrix indices in ${\mathfrak a}_i$ and $A_\mu$ will not preserve gauge invariance. 

\subsection{Dequantising  spin}
\label{SUBSEC:wong}

In \cite{stone-dwivedi} we resolved the problem of gauge invariance  by   {\it de-quantizing\/}   the gauge group representation  matrices  by introducing an   internal group-valued degree  of freedom   $g(t)\in G$ whose classical mechanics, when  {\it re-quantized\/}  by using the  geometric quantization procedure  of Kostant, Kirillov and Souriau \cite{kirillov}  would give us back the  matrices.   How to do this  is explained in physicist's language in  
\cite{wong, balachandran} and also in \cite{stone-dwivedi}.  Briefly stated, a   representation of $G$ with highest weight ${\Lambda}$
determines  a Lie algebra   element $\alpha_\Lambda $  that lies in the    Cartan sub-algebra.  
 The  quantum representation Hilbert space is  replaced by a classical phase space ${\mathcal O}_\Lambda$  that is the adjoint orbit  of $\alpha_\Lambda$. This orbit can be identified with the coset $G/H$, where the isotropy group $H$ is the subgroup of $G$ whose adjoint  action $\alpha_\Lambda \mapsto g \alpha_\Lambda g^{-1}$  leaves $\alpha_\Lambda$ fixed. 
The matrix-valued gauge field components $A_\mu$ and $F_{\mu\nu}$ are  replaced by   functions ${\mathcal O}_\Lambda\to  {\mathbb  R}$ as 
\bea
A_\mu \mapsto \widetilde A_\mu&\stackrel{\rm def}{=}&\tr(QA_\mu), \nonumber\\
F_{\mu\nu}\mapsto \widetilde F_{\mu\nu} &\stackrel{\rm def}{=} &\tr(QF_{\mu\nu}),
\eea
where 
\be
Q {=} g\alpha_\Lambda g^{-1}=Q^a\lambda_a.
\ee
Here  the trace is taken in some fixed faithful representation (most conveniently the defining representation of $G$).
We can use this  fixed trace to define a metric
\be
\gamma_{ab}= \tr\{\lambda_a\lambda_b\}
\label{EQ:lie-metric}
\ee
that we will use to raise or lower indices on Lie-algebra tensors such as the structure constants ${f_{ab}}^c$. In particular    $Q_a = \gamma_{ab}Q^c$ is the classical analogue of the generator $\lambda_a$, and the commutation relations (\ref{EQ:liealgebra}) are replaced by classical Poisson-bracket relations
\be
\{Q_a,Q_b\}= i{f_{ab}^c} Q_c.
\ee

The obvious way  to treat the non-Abelian Berry connection  is to repeat this scheme.  We therefore introduce  $\sigma(t)\in {\rm Spin}(2N)$ to acommodate  the  spin dynamics.  
We then  have an element  $\beta_{\mathfrak s}\in {\rm Cartan}({\rm Spin}(2N))$  that is determined by the particle's spin $\mathfrak s$ (see appendix \ref{SEC:trace-integrals} for details), an  orbit   
${\mathcal O}_{\mathfrak s}$, and the substitution 
\bea
{\mathfrak a}_i\mapsto \widetilde{\mathfrak a}_i &\stackrel{\rm def}{=}&\tr(\mathfrak S {\mathfrak a}_i),  \nonumber\\
{\mathfrak F}_{ij}\mapsto \widetilde {\mathfrak F}_{ij}&\stackrel{\rm def}{=}&\tr({\mathfrak S} {\mathfrak F}_{ij}),
\eea
where
 \be
  {\mathfrak S}\stackrel{\rm def}{=}\sigma\beta_{\mathfrak s} \sigma^{-1}= {\mathfrak S}^{(m)} X_{(m)}.
  \label{EQ:beta-s}
\ee   

\subsection{The   action and the equations of motion}
\label{SUBSEC:wong-action}

We  now assemble these  ingredients to write down the natural candidate for the   classical  action that   describes the motion of  a  Weyl fermion in a  background  non-Abelian gauge field in 2N+2 space-time dimensions:
\be
 S[\x,\p, g,\sigma]=\int \! dt\!\left(p^i \dot x^i  -|p| +i\tr\!\!\left\{\alpha_\Lambda g^{-1}\!\! \left (\frac{d}{dt}  -i(A_0+\dot x^i A_i) \right)g\right\}
- i\tr\!\!\left\{\beta_{\mathfrak s} \sigma^{-1}\!\!\left(\frac{d}{dt}  -i\dot x^i {\mathfrak a}_i \right)\sigma\right\}\right).
 \label{EQ:non-Abelian-action}
 \ee
 The proposed  action is  clearly  invariant under the gauge transformation 
\bea
g(t)&\to& h^{-1}(x(t),t)g(t),\nonumber\\
A_\mu&\to&  h^{-1}A_\mu h +ih^{-1}\frac{\partial}{\partial x^\mu} h.
\label{EQ:A-gauge}
\eea
It  is also invariant  under a ${\bf p}$-dependent change of basis $\ket{{{\bf p},\beta,+}}\to \ket{{{\bf p},\alpha,+}}U_{\alpha\beta}({\bf p})$ that takes 
\bea
\sigma(t)&\to& U^{-1}({\bf p})\sigma(t),\nonumber\\
{\mathfrak a}_j&\to& U^{-1}{\mathfrak a}_jU+iU^{-1}\frac{\partial}{\partial p^j} U.
\label{EQ:a-gauge}
\eea

The equations of motion 
that arise from varying $g\to g(1+g^{-1} \delta g)$ and $\sigma\to \sigma(1+\sigma^{-1}\delta \sigma)$ are
\bea
[\alpha_\Lambda, g^{-1}(\partial_t -i(A_0+\dot x^i A_i )g]&=&0,\nonumber\\
{}[\beta_{\mathfrak s}, \sigma^{-1}(\partial_t -i\dot p^i {\mathfrak a}_i)\sigma]&=&0.
\label{EQ:ambiguous}
\eea
These equations do not uniquely determine $\dot g$ and $\dot \sigma$. Indeed any solution   $g(t)$ and $\sigma(t)$  can be multiplied on the right by an arbitrary
 time-dependent  element of the corresponding isotropy subgroup and still satisfy the equation.  As a result $g$ and $\sigma$  must  be thought of as living in the cosets ${\mathcal O}_\Lambda$ and ${\mathcal O}_{\mathfrak s}$.    The  time evolution of the Lie-algebra-valued quantities  $Q$ and ${\mathfrak S}$  is insensitive to the ambiguity, and from (\ref{EQ:ambiguous}) we find that
\bea
\dot Q&=&-i[Q, A_0+ \dot x^i A_i],\nonumber\\
\dot {\mathfrak S}&=&-i[{\mathfrak S}, {\mathfrak a}_i\dot p^i].
\label{EQ:dotQdotSigma0}
\eea 
In components these equations read
\bea
\dot Q^c &=& {f_{ab}}^c Q^a  (A^b_0+ \dot x^i A^b_i),\nonumber\\
\dot {\mathfrak S}^{(c)}&=&{f_{(a)(b)}}^{(c)} {\mathfrak S}^{(a)} {\mathfrak a}_i^{(b)}\dot p^i.
\label{EQ:dotQdotSigma}
\eea 
 
The remaining equations of motion 
\bea
\dot p^i&=& \tr\{Q(F_{i0}+F_{ij}\dot x^j)\},\nonumber\\
\dot x^i &=& \hat p^i - \tr\{{\mathfrak S} {\mathfrak F}_{ij}\}\dot p^j.
\label{EQ:xpEofM}
\eea
are more straightforward, although in principal we still have to solve them for $\dot x^i$ and $\dot p^i$ in terms of the other degrees of freedom as we did in \cite{stone-dwivedi}. This task is easy in three  space  dimensions, but complicated in dimensions higher than three. For this reason, in subsequent sections we develop tools that allow us to deduce the consequences of  (\ref{EQ:xpEofM})  without finding explicit expressions for $\dot x^i$ and $\dot p^i$.

 \section{A generalized  Liouville theorem}
 \label{SEC:liouville} 
 
 Both the gauge and abelian chiral  anomalies  can be understood physically as  a  spectral flow of states from the  infinitely deep negative-energy Dirac sea, through the diabolical \cite{berry-wilkinson}  Dirac point, and into  the  finite-depth positive-energy Fermi sea. If we monitor    the phase-space  density only in the positive energy  region, the influx of states will appear as a source term located at ${\bf p}=0$ in the Liouville phase-space volume conservation law. 
We therefore  begin by recalling the proof of  Liouville's theorem,  and how the theorem  is  modified   when the symplectic structure is allowed to depend on time.

\subsection{Extended phase space}
\label{SUBSEC:extended}

Consider a general even-dimensional  phase space $M$ with co-ordinates ${\bm \xi}= (\xi^1,\ldots, \xi^{2n})$ and equipped with a Hamiltonian  action functional 
\be
S[\bm \xi]=\int dt\left\{\sum_{i=1}^{2n}\eta_i({\bm\xi},t) \dot \xi^i-H({\bm \xi},t)\right\}.
\label{EQ:general-form-action}
\ee
Demanding  that  $\delta S=0$ under  a  variation $ \xi^i \to \xi^i+\delta \xi^i$    results in the equations of  motion 
\be
\left(\frac{\partial \eta_j}{\partial \xi^i}- \frac{\partial \eta_i}{\partial\xi^j}\right)\dot \xi^j=\left(\frac{\partial H}{\partial \xi^i}+\frac{\partial \eta_i}{\partial t}\right).
\label{EQ:EofM}
\ee
We will use the notation
\be
\omega_{ij}=\frac{\partial \eta_j}{\partial \xi^i}- \frac{\partial \eta_i}{\partial\xi^j}.
\ee
For the  slightly generalized set of Hamilton's  equations (\ref{EQ:EofM}) to be solvable for $\dot \xi^i$ without constraints, the symplectic matrix $\omega_{ij}$ must be invertible at every point in $M$. We assume that this condition is satisfied.
Now, from  (\ref{EQ:EofM})  and   the antisymmetry of $\omega_{ij}$,  we see that the  $\dot \xi^i$ automatically satisfy the condition
\be
\dot \xi^i\left(\frac{\partial H}{\partial \xi^i}+\frac{\partial\eta_i}{\partial t}\right)=0.
\label{EQ:EofMconsequence}
\ee

As with the usual  Hamiltonian formalism, many results are most compactly obtained with  the tools  of vector fields and differential forms. To use these  in the time dependent setting   it is convenient to  extend the even-dimensional phase space $M$ to the odd-dimensional space $M'=M\times {\mathbb R}$, where ${\mathbb R}$ is the time co-ordinate.  
 We may then combine the $\eta_i$ and the hamiltonian $H$ into  a  one-form
 \be
\eta_H=\sum_{i=1}^{2n} \eta_id\xi^i -Hdt,  
\ee
 on $T^*M'$.    Let   $\omega_H=d\eta_H$  be  its  exterior derivative
\be
\omega_H= \frac 12  \omega_{ij}d\xi^i d\xi^j-\left(\frac{\partial \eta_i}{\partial t} +\frac{\partial H}{\partial \xi^i}\right) d\xi^i\,dt.
\ee

Define a vector field 
\be
{\bf v}=\frac{\partial}{\partial t} +\dot \xi^j \frac{\partial}{\partial \xi^j} .
\label{EQ:hamiltonian-v}
\ee
and take the  interior product of ${\bf v}$ with the two-form $\omega_H$  to get
\be
i_{\bf v}\omega_H=\left(-\omega_{ij}\dot \xi^j+\frac{\partial H}{\partial\xi^i}+\frac{\partial \eta_i}{\partial t}\right)d\xi^i -\dot \xi^i\left(\frac{\partial H}{\partial \xi^i}+\frac{\partial \eta_i}{\partial t} \right)dt.
\ee
From (\ref{EQ:EofM}) and (\ref{EQ:EofMconsequence}) we see that  the compact equation 
\be
 i_{\bf v}\omega_H=0
 \ee
 is completely equivalent to the equations of motion. 

Any  odd-dimensional two-form  such as $\omega_H$ must  possess at least one null vector. Here, the  matrix $\omega_{ij}$ being invertible ensures that  the null  space of $\omega_H$ is {\it precisely\/}   one dimensional.  Collectively the   null spaces compose  the {\it characteristic bundle\/} over $M'$ of the {\it  contact structure\/}  $\omega_M$\cite{abrahams-marsden}, and   the ${\bf v}({\bm \xi},t)$ determined by  the equations of motion is the unique vector in the fibre over $({\bm \xi},t)$ that has   unity as the coefficient of $\partial/\partial t$.

\subsection{The Liouville measure}

We  can  define  a volume ($2n+1$)-form on the extended phase space by
\bea
\Omega &\stackrel{\rm def}{=}& \frac{1}{n!}\omega_H^ndt,\nonumber\\
&=& \frac{1}{n!}\omega^ndt.
\label{EQ:vol_definition}
\eea
Here  
\be
\omega= \frac 12  \omega_{ij}d\xi^i d\xi^j,
\ee
is the usual phase space symplectic form, now allowed to be time dependent. The  second line of (\ref{EQ:vol_definition}) follows from the first because the explicit factor of $dt$ excludes all  $d\xi^idt$ terms in $\omega^n_H$.

We can use the first line of (\ref{EQ:vol_definition})  to compute  Lie derivative of $\Omega$ with respect to ${\bf v}$. We find  
\bea
{\mathcal L}_{\bf v} \Omega&=& (i_{\bf v}  d+di_{\bf v})\Omega,\nonumber\\
&=& di_{\bf v}\Omega,\,\,\,\quad \qquad \qquad \qquad \hbox{($d\Omega=0$ because  $\Omega$  is a top form)}\nonumber\\
&=& \frac 1 {n!} d((i_{\bf v} \omega_H^n)dt+\omega_H^n), \quad\hbox{($i_{\bf v}$ is an antiderivation, and $\omega_H^n$ is even)}\nonumber\\
&=& \frac 1 {n!} d(\omega_H^n),\,\qquad\qquad\qquad\hbox{($i_{\bf v}\omega_H^n\equiv n\omega_H^{n-1}\wedge  i_{\bf v}\omega_H=0$ by the equations of motion)}\nonumber\\
&=&0. \,\,\quad \qquad\qquad\qquad\qquad \hbox{($d^2\eta_H=0$ provided $\eta_H$ is nowhere-singular)}
\label{EQ:form-liouville}
\eea

We can alternatively compute the Lie derivative of $\Omega$ from  the  second line of (\ref{EQ:vol_definition}). We observe that   
\be
\frac{1}{n!} \omega^n dt= \sqrt{\omega}d\xi^1\cdots d\xi^{2n}dt,
\ee
where $\sqrt{\omega}\equiv \sqrt{{\rm det}(\omega)}$ is the Pfaffian ${\rm Pf}(\omega)$ of the skew-symmetric matrix $\omega_{ij}$.
From   the   derivation property of the Lie derivative we now  find 
\bea
{\mathcal L}_V \Omega &=&
\left({\mathcal L}_{\bf v}\sqrt{\omega}\right)d\xi^1\cdots d\xi^{2n}dt+ \sqrt{\omega}\left({\mathcal L}_v d\xi^1\cdots d\xi^{2n}dt\right)\nonumber\\
&=&\left(\frac{\partial \sqrt{\omega}}{\partial t}+ \dot \xi^i\frac{\partial \sqrt{\omega}}{\partial \xi^i}\right)d\xi^1\cdots d\xi^{2n}dt + \sqrt{\omega}\left(\frac{\partial \dot \xi^i}{\partial \xi^i} d\xi^1\cdots d\xi^{2n}dt\right)\nonumber\\
&=&\left\{\frac{\partial \sqrt{\omega}}{\partial t}+ \frac{\partial \sqrt{\omega}\dot \xi^i}{\partial \xi^i}\right\}d\xi^1\cdots d\xi^{2n}dt.
\eea
Comparing  the  two computations   shows that our    generalized Hamilton equations lead to 
\be
\frac{\partial \sqrt{\omega}}{\partial t}+ \frac{\partial \sqrt{\omega}\dot \xi^i}{\partial \xi^i}=0.
\label{EQ:t-dep-liouville}
\ee
This last equation  is the time-dependent version  of  Liouville's theorem.  In our application, the manipulations in   (\ref{EQ:form-liouville}) will fail at the last step  because a  generalization of the  Berry-phase monopole leads    $\omega_H^n$ to  be singular. Consequently  (\ref{EQ:t-dep-liouville}) will be violated by  a source term  at   the Dirac point.   
The remaining sections of this paper will be devoted to finding  the strength of   this source term in terms of the  external fields acting on our particles.

\section{The Boltzmann equation} 
\label{SEC:boltzmann}

Now we apply the formalism of section \ref{SEC:liouville} to the  action functional (\ref{EQ:non-Abelian-action}).  The extended phase space we need is $M' ={\mathbb R}^{2N+2}\times   {\mathcal O}_\Lambda\times {\mathcal O}_{\mathfrak s}\times {\mathbb R} $ with ${\mathbb R}^{2N+2}$ being the particle's  $(\x,\p)$ co-ordinates, ${\mathcal O}_\Lambda$, ${\mathcal O}_{\mathfrak s}$ being  the internal gauge and spin spaces, and ${\mathbb R}$ time. We will  avoid    as much as possible the use  explicit co-ordinates $\xi^i$ on the internal  spaces, and instead use  intrinsic geometric quantities.

\subsection{Boltzmann and Liouville}

We begin by writing  (\ref{EQ:non-Abelian-action})  as an integral along the phase-space trajectory.   We  need  to define the    differential forms 
 \be
 A= A_idx^i+A_0dt,\quad {\mathfrak a}= {\mathfrak a}_i dp^i,
 \ee
 as well as  
 \be
 \omega_R^Q=dg g^{-1}, \quad \omega_R^{\mathfrak S}=d\sigma \sigma^{-1},
 \ee 
 which are pullbacks to the  trajectory in $(\x,\p, g, \sigma)$ space of the right-invariant Maurer-Cartan forms  on $G$ and on ${\rm Spin}(2N)$ respectively. 
 The action (\ref{EQ:non-Abelian-action}) then becomes  
 \bea
S[\x,\p,g,\sigma] &=& \int \left(p^i dx^i -|p|dt +\tr\{Q(i\,dgg^{-1} +A)\} -\tr\{{\mathfrak S} (i\,d\sigma \sigma^{-1} +{\mathfrak a})\}\right)\nonumber\\
 &=&  \int \left(p^i dx^i -|p|dt +\tr\{Q(i \omega_R^Q+A)\} -\tr\{{\mathfrak S} (i \omega_R^{\mathfrak S} +{\mathfrak a})\}\right).
 \label{EQ:non-Abelian-action-form}
 \eea
Now  (\ref{EQ:non-Abelian-action-form}) is of the general form
(\ref{EQ:general-form-action}) with 
\be
\eta_H=p^i dx^i -|p|dt +\tr\{Q(i \omega_R^Q+A)\} -\tr\{{\mathfrak S} (i \omega_R^{\mathfrak S} +{\mathfrak a})\}.
\ee
Using 
$F=dA-iA^2$, ${\mathfrak F}=d{\mathfrak a}-i{\mathfrak a}^2$ and  $d\omega_R= (\omega_R)^2$ we find that
\be
\omega_H= dp^idx^i -d|p| dt +\widetilde F -\widetilde{\mathfrak F} -i \tr\{Q(\omega_R^Q-iA)^2\} +i \tr\{{\mathfrak S}(\omega_R^{\mathfrak S}-i{\mathfrak a})^2\},
\ee
and the  volume  form (\ref{EQ:vol_definition})  becomes  
 \be
 \Omega= \frac{1}{M!} \omega_H^M dt,
 \label{EQ:ourliuoville}
 \ee
 where   $M= (2N+1)+ m_\Lambda +m_{\mathfrak s}$ with   $m_\Lambda ={\rm dim}({\mathcal O}_\Lambda)/2$, $m_{\mathfrak s}={\rm dim}({\mathcal O}_{\mathfrak s})/2$.

 There are potentially many terms in the high power of $\omega_H$ appearing in (\ref{EQ:ourliuoville}). A considerable  simplification arises, however, 
because  all terms involving  explicit $A$'s  and $\mathfrak a$'s must cancel.
  This cancellation can be tediously verified in 3+1 dimensions, but can be shown  to occur  in general  by observing that at any chosen point in $\x,\p$ space we can make  gauge transformations so that both $A$ and $\mathfrak a$ (but not their derivatives) vanish.  The  transformations  (\ref{EQ:A-gauge}) and (\ref{EQ:a-gauge})
 that return us to the original gauge will leave this $\Omega({\mathfrak a}=A=0)$ volume form unchanged ---  either because  traces are invariant under adjoint actions, or because the inhomogeneous  $g^{-1}dg$ and $\sigma^{-1}d\sigma$  terms   will seek  to introduce extra $dx^\mu$'s,  $dt$'s, or $dp^i$'s  
 that  are forbidden by antisymmetry. 
  
  Thus we find that    
\be
 \frac{1}{M!} \omega_H^M=    \frac{1}{(2N+1)!}(dp^idx^i  +\widetilde F -\widetilde {\mathfrak F})^{2N+1} d\mu_\Lambda d\mu_{\mathfrak s},
 \ee
 where
 \be
 d\mu_\Lambda= \frac{1}{m_\Lambda!}[ i \tr\{\alpha_\Lambda (\omega_L^Q)^2\} ]^{m_\Lambda},\qquad d\mu_{\mathfrak s}=  \frac{1}{m_{\mathfrak s}!}[- i \tr\{\beta_{\mathfrak s}(\omega_L^{\mathfrak S})^2)\}]^{m_{\mathfrak s}}, 
\ee
are, up to factors, the  Kirillov-Kostant measures on the co-adjoint orbits ${\mathcal O}_\Lambda$ and ${\mathcal O}_{\mathfrak s}$ respectively.
We have taken advantage of the absence of the gauge fields to introduce the left-invariant Maurer Cartan forms $\omega_L^Q= g^{-1} dg$ and $\omega_L^{\mathfrak S}= \sigma^{-1}d\sigma$.

We also find that 
\be
{\bf v}= \frac{\partial}{\partial t} + \dot x^i \frac{\partial}{\partial x^i}+ \dot p^i \frac{\partial}{\partial x^i} + (g^{-1}\dot g)^a L^{\rm gauge} _a+ (\sigma^{-1} \dot \sigma)^{(a)} L^{\rm spin}_{(a)}
\ee
is the appropriate form  for the  Hamiltonian  vector field ${\bf v}$   of equation   (\ref{EQ:hamiltonian-v}).  Here  $ L^{\rm gauge} _a$ is the  left-invariant vector field
dual to the left-invariant Maurer-Cartan form. We have 
\be
  \omega_L^Q(L^{\rm gauge}_a)= \lambda_a,
\ee
so  $\omega_L^Q({\bf v})= (g^{-1}\dot g)^a\lambda_a= g^{-1}\dot g$ is the analogue of  $dx^i({\bf v}) =\dot x^i$.
Similarly   $L^{\rm spin}_{(a)}$  is dual to the ${\rm Spin}(2N)$ Maurer-Cartan form: 
\be
\omega^{\mathfrak S}_L (L^{\rm spin} _{(a)})=  X_{(a)}
\ee

In order  to write down the Liouville theorem we will need to know how to compute the 
Lie derivative of the adjoint orbit measures. After a  little effort  and the use of (\ref{EQ:dotQdotSigma})  we find
\be
{\mathcal L}_{\bf v} d\mu_\Lambda = {f_{ab}}^c Q^a A^b_i \frac{\partial \dot x^i}{\partial Q^c}d\mu_\Lambda+\ldots,
\ee
where the dots indicate terms involving $dx^i$ or $dt$ that have no effect in  $d\mu_\Lambda dx^1\cdots dx^{2N+1}dt$.
There is   an analogous  expression for the Lie derivative of $\mu_{\mathfrak s}$.

We now define $\sqrt{\omega}$ by  
\be
\Omega= \sqrt{\omega}\, d\mu_\Lambda d\mu_{\mathfrak s}  dp^1\cdots dp^{2N+1} dx^1\cdots dx^{2N+1} dt.
\ee
If---for the duration of this section only---we ignore the singularity at $\p=0$  and follow the procedure  in section \ref{SEC:liouville} we end up with with ${\mathcal L}_{\bf v}\Omega=0$ being equivalent to 
\be
\nabla_t \sqrt{\omega}+ \nabla_{x^i} \sqrt\omega \dot x^i +\nabla_{p^i} \sqrt \omega \dot p^i=0,
\ee
where
\bea
\nabla_t &=& \frac{\partial}{\partial t}+ {f_{ab}}^c A^a_0 Q^b  \frac{\partial }{\partial Q^c}, \nonumber\\
\nabla_{x^i}&=& \frac{\partial} {\partial x^i}+ {f_{ab}}^c  Q^a A^b_i\frac{\partial }{\partial Q^c}, \nonumber\\
\nabla_{p^i} &=& \frac{\partial} {\partial p^i}
+{f_{(a)(b)}}^{(c)} a_k^{(a)}{\mathfrak S}^{(b)} \frac{\partial }{\partial {\mathfrak S}^{(c)}}, 
\label{EQ:covariant}
\eea
are ``covariant derivatives''  that ensure invariance under gauge transformations.

We   introduce a phase-space density $f(\x,\p, Q, {\mathfrak S},t)$  and  let it be  advected with the flow 
\be
\left(\frac{\partial}{\partial t} +\dot x^i \frac{\partial}{\partial x^i} +\dot p^i \frac{\partial}{\partial p^i}+ \dot Q^a \frac{\partial}{\partial Q^a}+\dot {\mathfrak S}^{(a)} \frac{\partial}{\partial {\mathfrak S}^{(a)}}\right)f=0.
\label{EQ:boltzmann}
\ee
This  advection condition is the collisionless   single-particle Boltzmann equation for our system. It generalizes the 3+1 dimensional Boltzmann equation in \cite{litim}.

We can use (\ref{EQ:dotQdotSigma})  and (\ref{EQ:covariant})  to group these terms as
\be
\left(\nabla_t +\dot x^i \nabla_{x^i} +\dot p^i \nabla_{p^i}\right)f=0, 
\ee
and so find that  
 \be
{\mathcal L}_{\bf v}( f\Omega)=0,
\label{EQ:allLiezero}
\ee
which expresses the conservation of  probability.

\subsection{Conservation laws}
\label{SUBSEC:conservation}

The continuity equation  (\ref {EQ:allLiezero}) for the phase space density is  the origin of  various  conservation laws. 
From  (\ref {EQ:allLiezero}) we can derive the conservation 
\be
\frac{\partial J^0}{\partial t}+\frac{\partial J^i}{\partial x^i}=0
\label{EQ:abelian-conservation}
\ee
of the 
particle-number current
\bea
J^{0}(t,{\x})&=& \int f(\x,\p, Q, {\mathfrak S},t)  \sqrt{\omega}  \,d\bar \mu_\Lambda\bar d \mu_{\mathfrak s}  \frac{d^{2N+1}p}{(2\pi)^{2N+1}} ,\nonumber\\ 
J^{i}(t,{ \x}) &=& \int    \dot x^i f(\x,\p, Q, {\mathfrak S},t) \sqrt{\omega} \, d\bar \mu_\Lambda\bar d\mu_{\mathfrak s}  \frac{d^{2N+1}p}{(2\pi)^{2N+1}},
\eea
and the covariant conservation
\be
\frac{\partial J^0_a}{\partial t} -{f_{ab}}^c A^b_0 J^0_c +\frac{\partial J^i_a}{\partial x^i} -{f_{ab}}^c A^b_i J^i_c=0
\label{EQ:nonabelian-conservation}
\ee
of the gauge current
\bea
J^{0}_a(t,{\x})&=& \int Q_a f(\x,\p, Q, {\mathfrak S},t)   \sqrt{\omega} \, \bar  d\mu_\Lambda d\bar \mu_{\mathfrak s} \frac{d^{2N+1}p}{(2\pi)^{2N+1}} ,\nonumber\\ 
J^{i}_a(t,{\x}) &=& \int   Q_a \dot x^i f(\x,\p, Q, {\mathfrak S},t) \sqrt{\omega} \, \bar d \mu_\Lambda d\bar \mu_{\mathfrak s}\frac{d^{2N+1}p}{(2\pi)^{2N+1}}.
\eea
In each of these definitions  the integral is over all of momentum space and over both  adjoint orbits ${\mathcal O}_\Lambda$,  ${\mathcal O}_{\mathfrak s}$.
While  writing down the expressions for the currents we   have  taken the opportunity to normalize the measure factors. We have put a $1/2\pi$ with each $dp$ and rescaled the adjoint orbit measure so that 
\be
d\bar  \mu_\Lambda= \frac{1}{(2\pi)^{m_\Lambda} m_\Lambda!}  [i\,\tr\{\alpha_\Lambda (\omega_L^Q)^2\} ]^{m_\Lambda},\qquad d\bar \mu_{\mathfrak s}=  \frac{1}{(2\pi)^{m_{\mathfrak s}}m_{\mathfrak s}!}[i\, \tr\{\beta_{\mathfrak s}(\omega_L^{\mathfrak S})^2)\}]^{m_{\mathfrak s}}, 
\ee
This rescaling is one place where we need knowledge of quantum mechanics:  the normalized Liouville measure  counts  (approximately) one quantum state per  unit volume of the classical phase space. Consequently  the exclusion principle says that the maximum allowed value of $f(x,p, Q, {\mathfrak S},t)$ is unity.

To see that (\ref{EQ:allLiezero}) leads to the conservation laws (\ref{EQ:abelian-conservation}) and (\ref{EQ:nonabelian-conservation})  recall that if we have a $q$-dimensional manifold $M$ and  integrate a  $p$-form $\theta$ over a smooth  $p$-dimensional region $N(\tau)$, each point of which  is being advected by a flow ${\bf v}= d{\bf x}/{d\tau}$,  then Liebniz'  integral formula reads  
\be
\frac{d}{d\tau} \int_{N(\tau)}\theta = \int_{N(\tau)}{\mathcal L}_{\bf v} \theta. 
\label{EQ:liebniz}
 \ee
 An immediate   corollary  is that when we   integrate a q-form $\Theta$ over the entirety of $M$ we have
 \be
 \int_M {\mathcal L}_ {\bf v} \Theta=0.
 \label{EQ:corollary}
 \ee
 Now  (\ref{EQ:abelian-conservation}) is equivalent to the vanishing of 
 \be
 \int_{{\mathbb R}^{2N+1}\times {\mathbb R}} \varphi(\x,t) \left( \frac{\partial J^0}{\partial t}+\frac{\partial J^i}{\partial x^i}\right) d^{2N+1} x\, dt
 \ee
 for any test function $\varphi(\x,t)$. But ${\mathcal  L}_{\bf v} (f\Omega)=0$ and (\ref{EQ:corollary}) gives us 
 \bea 
 0&=& \int_{M'}\mathcal L_{\bf v} ( \varphi f\Omega)\nonumber\\
 &=& \int_{M'} \left( \frac{\partial \varphi }{\partial t}+\dot x^i \frac{\partial \varphi }{\partial x^i}\right) f\Omega\nonumber\\
 &=& -   \int_{{\mathbb R}^{2N+1}\times {\mathbb R}} \varphi(x,t) \left( \frac{\partial J^0}{\partial t}+\frac{\partial J^i}{\partial x^i}\right)  d^{2N+1} x\, dt.
 \eea
 A similar calculation gives (\ref{EQ:nonabelian-conservation}).

\section{The Liouville anomaly} 
\label{SEC:anomalies}

The extended phase space differential-form language introduced in the previous section may seem rather abstruse, but a glance at 
appendix \ref{SEC:anomaly-3d}  will  reveal that  it  provides a more compact and  transparent   derivation of the Abelian anomaly than that in \cite{stephanov, stone-dwivedi}. In this section 
 we will generalize the derivation in the appendix by  identifying  the Liouville-theorem source term in  any even space-time  dimension and use it to derive the anomalous conservation laws.

We begin with an observation about the Liouville measure. In  
\be
 \frac{1}{(2N+1)!}(dp^idx^i  +\widetilde F -\widetilde {\mathfrak F})^{2N+1} dt = \sqrt{\omega} dp^1\cdots dp^{2N+1} dx^1\cdots dx^{2N+1} dt
\ee
the measure factor $\sqrt \omega$ is the Pfaffian of the  $(4N+2)$-by-$(4N+2)$ matrix
\be
\omega_{ij}= \left(\matrix{ -\widetilde {\mathfrak F}  &{\mathbb I}\cr -{\mathbb I}&\widetilde F}\right)_{ij}.
\ee
The factor $dt$ excludes $\widetilde F_{0\mu}$ from being an entry in the block submatrix ${\widetilde F}$ appearing here, so it has the same number of rows and columns as does $\widetilde {\mathfrak F}$.
The Schur determinant formula now shows that 
\be
\det\left(\matrix{ -\widetilde {\mathfrak F}  &{\mathbb I}\cr -{\mathbb I}&\widetilde F}\right) = \det({\mathbb I}-{\widetilde F \widetilde{\mathfrak F}}),
\ee
and the  Liouville measure is a square-root of this determinant.
This root  can be expressed in terms of a    ``double Pfaffian''
\be
 {\rm Pf}(\widetilde {\mathfrak F}, \widetilde F)\stackrel{\rm def}{=}  \sum_{k=0}^{N}\sum_{I_{2k}} {\rm Pf}(\widetilde {\mathfrak F}_{I_{2k}}). {\rm Pf}({ \widetilde F}_{I_{2k}}),
\ee
where    each $I_{2k}$ is a cardinality-$2k$   subset of the indices on the matrices, and the term with $k=0$ is understood to be unity.
The square root is ambiguous, and we  actually have 
\be
\sqrt{\omega}= \pm  {\rm Pf}(\widetilde {\mathfrak F}, \widetilde F), 
\ee
where the  $\pm$ sign depends on how many $dx^i$'s  and $dp^i$'s  have to be interchanged to turn $(dx^idp^i)^{2N+1}$ into   $d^{2N+1}p \,d^{2N+1}x$.  This sign is unimportant provided we use a  consistent  definition in   the   current and anomaly measures.

  When we expand out $\omega_H^M/M!$ there is no longer an {\it explicit\/}  $dt$ and so  $\widetilde F_{0\mu}$ is allowed. In particular one of the many terms that appear in $\omega_h^M/M!$ is  
  \be
\frac{(-1)^N}{N!} \left(\frac{\widetilde {\mathfrak F}}{2\pi}\right) ^N\   \frac{1}{(N+1)!}\left(\frac{\widetilde F}{2\pi}\right)^{N+1}d\bar\mu_{\Lambda}d\bar \mu_{\mathfrak s}
   \label{EQ:badterm}
 \ee
 where the $2\pi$'s come from the factors of $1/2\pi$ that we put with each $dp^i$.

 It is this term that causes $d\omega_H^M/M!\ne0$ in the last line of (\ref{EQ:form-liouville}) and hence  gives rise to  the anomaly.  We show in the appendix that the integral 
 of the Chern character  
 \be 
 \frac{1}{N!(2\pi )^N} \int_{S}  \tr({\mathfrak F} ^N) 
 \ee
of the spin-connection curvature 
over a closed 2N-surface $S$  in momentum space is $(-1)^N$ (for positive  helicity)  when  $S$ encloses ${\bf p}=0$, and is zero otherwise.    
As with the Abelian monopole, we can conveniently abbreviate this statement as 
\be
 d\left( \frac{1}{N!(2\pi )^N}  \tr\left({\mathfrak F}^N\right)\right)=(-1)^N  \delta^{2N+1}({\bf p})dp^1\cdots dp^{2N+1},
 \label{EQ:chern-monopole}
 \ee 
 exhibiting a higher-dimensional version of (\ref{EQ:monopole}).  
For the gauge field, however,  we have 
 \be
 d\left(\frac{1}{(N+1)!}\tr\left(\frac{F}{2\pi}\right)^{N+1}\right)=0
 \label{EQ:closed}
 \ee
 by well known properties of characteristic classes.
 
 In (\ref{EQ:chern-monopole}) of course we have the matrix-valued curvature  ${\mathfrak F}$ and a trace over its  quantum indices. Similarly in (\ref{EQ:closed}). In (\ref{EQ:badterm}) we have the function-valued $\widetilde {\mathfrak F}$ and $\widetilde F$  and are to integrate over the adjoint orbits ${\mathcal O}_{\mathfrak s}$ and ${\mathcal O}_\Lambda$.  However we also show in the appendix that  provided we take integral  not over  the na{\"i}ve orbit associated with the greatest weight of the spin representation but instead over the orbit associated  with the {\it Weyl shifted\/} weight, then   the quantum and classical traces coincide.  Similarly the  classical trace ${\mathcal O}_\Lambda$ is always proportional to the quantum trace, and so (\ref{EQ:closed}) remains true when $F$ is replaced by $\widetilde F$. 
 We also show that (\ref{EQ:badterm})  is the only contribution to $d\omega_H^M/M!$  that survives the trace operation. 
 
 One additional ingredient is required.  It is not the integral of ${\mathcal L}_{\bf v}\Omega$ that is needed in the conservation law, but the integral of ${\mathcal L}_{\bf v}(f\Omega)$. If we are to get the standard expression for the anomaly we must have $f$ be identically unity over and within a surface sufficiently far from $p=0$ that the adiabatic approximation and the resultant classical machinery be applicable.  Thus we need a Fermi sea that is deep enough that finite temperature effects do not depopulate the sea too close to the Dirac point.    
 
 Assuming that $f$ is indeed unity in the region where it is needed, we immediately find that  contribution of the delta function (\ref{EQ:chern-monopole}) to ${\mathcal L}_{\bf v}\Omega$  modifies the conservations laws to read
 \be
\frac{\partial J^0}{\partial t}+\frac{\partial J^i}{\partial x^i}=\frac{ \epsilon^{i_1i_2 \ldots i_{2N+1} i_{2N+2}}}{(2\pi)^{N+1} 2^{N+1}(N+1)!}  \int _{{\mathcal O}_\Lambda} \left(\widetilde F_{i_1i_2}\cdots  \widetilde F_{i_{2N+1}i_{2N+2}}\right) d\bar \mu_\Lambda
\label{EQ:abelian-anomaly}
\ee
 and
 \be
\frac{\partial J^0_a}{\partial t} -{f_{ab}}^c A^b_0 J^0_c +\frac{\partial J^i_a}{\partial x^i} -{f_{ab}}^c A^b_i J^i_c=
\frac{ \epsilon^{i_1i_2 \ldots i_{2N+1}i_{2N+2}}}{(2\pi)^{N+1} 2^{N+1}(N+1)!}  \int _{{\mathcal O}_\Lambda} Q_a \left(\widetilde F_{i_1i_2}\cdots  \widetilde F_{i_{2N+1}i_{2N+2}}\right) d\bar \mu_\Lambda 
\label{EQ:nonabelian-anomaly}
\ee
The  phase space integrals are classical approximations to the symmetrized traces
\be
 \int _{{\mathcal O}_\Lambda} \left(\widetilde F_{i_1i_2}\cdots  \widetilde F_{i_{2N+1}i_{2N+2}}\right) d\bar \mu_\Lambda 
\sim \,{\rm str}_\Lambda\! \left\{F_{i_1i_2}\cdots F_{i_{2N+1}i_{2N+2}}\  \right\}
\ee
and
\be
 \int _{{\mathcal O}_\Lambda} Q_a \left(\widetilde F_{i_1i_2}\cdots  \widetilde F_{i_{2N+1}i_{2N+2}}\right) d\bar \mu_\Lambda 
\sim \,{\rm str}_\Lambda\! \left\{\lambda_a F_{i_1i_2}\cdots F_{i_{2N+1}i_{2N+2}}\  \right\}
\ee
taken in the representation with greatest weight vector $\Lambda$. With this substitution and setting $2N+2=2n$ equation (\ref{EQ:abelian-anomaly})   is the familiar expression for the singlet anomaly in $2n$ space-time dimensions
\be
\partial_\mu J^\mu =\frac{1}{(4\pi)^n n!} \epsilon^{\nu_1\ldots \nu_{2n}}{\rm tr}_{\Lambda}\{\hat \lambda_{a_1}\cdots \hat \lambda_{a_n}\} F^{a_1}_{\nu_1\nu_2}\cdots F^{a_n}_{\nu_{n-1}\nu_n},
\ee
and equation 
(\ref{EQ:nonabelian-anomaly}) is the covariant form 
\be
D_\mu J^\mu_a =\frac{1}{(4\pi)^n n!} \epsilon^{\nu_1\ldots \nu_{2n}}{\rm tr}_{\Lambda}\{\hat \lambda_a \hat \lambda_{b_1}\cdots \hat \lambda_{b_n} \}F^{b_1}_{\nu_1\nu_2}\cdots F^{b_n}_{\nu_{n-1}\nu_n},
\ee
of the non-Abelian gauge anomaly.  

 It  exhibited explicitly in \cite{stone-dwivedi} for  the case $G={\rm SU}(3)$, that the accuracy of the classical approximation to the symmetrized trace is again greatly increased if the classical trace integral is actually taken over the orbit corresponding to the Weyl-shifted weight $\Lambda\to \Lambda+\rho$, where  the Weyl vector $\rho$ is half the sum of the positive roots.  
 
\section{Discussion}

We have shown that the classical equations of motion derived from the single-particle action functional (\ref{EQ:non-Abelian-action}) give rise to the same anomalous conservation laws as  a Weyl fermion  in a background gauge field. That we get the correct form and coefficients in the anomalies gives strong evidence that  we correctly identified (\ref{EQ:non-Abelian-action})  as the appropriate generalization of the 3+1 abelian action (\ref{EQ:abelian-action}).  

The only quantum input into the equations of motion is the non-Abelian Berry connection that  performs  parallel transport of the degenerate helicity states.  On its own the Berry transport provides the {\it form} of the chiral anomaly. Additional quantum knowledge is required to obtain the correct {\it coefficient\/}. We  need to normalize the measure on the phase-space so that there is one quantum state per unit volume. For increased accuracy we should  also modify the phase space itself by Weyl-shifting the weight  that defines  the gauge charge of the particles,  and also the weight   that defines their spin.  

It is  interesting, but perhaps not surprising,  that  quantum Berry phase effects survive in the classical limit. The original example of the   abelian Berry phase is a spin  that is being forced to change its direction by a strong magnetic field. Here Berry's  gauge field ${\bf a}({\bf p})$ provides an analogue  Lorentz force that causes  Larmor precession. The gauge field is simply the way in which  the spin's gyroscopic stiffness manifests itself,  and so is best understood as a classical effect.  What our calculation shows is that  the quantum effects of the Fermi-surface Berry flux
\cite{haldane, son1,chen-pu-wang-wang} might also be best thought of as being the result of gyroscopic  forces.

 \section{Acknowledgements}   This  project was supported by the National Science Foundation  under grant  DMR 09-03291.

 \appendix 
 
\section{The Abelian  anomaly in 3+1 dimensions}
\label{SEC:anomaly-3d}
In this appendix we use the extended phase space formalism to obtain  the 
Abelian  anomaly in $3+1$ dimensions.   The reason for our doing so is to illustrate  how much more powerful  is this   formalism  compared to the method used in \cite{stephanov} and \cite{stone-dwivedi}.

The action is
\be
 S[{\bf x, p}] = \int  \eta_H = \int (p_i dx^i - |{\bf p}| dt + A - {\mathfrak a}) .
\ee
 Then  
\be
\omega_H = d\eta_H=  d(p_i dx^i - |{\bf p}| dt + A - {\mathfrak a}) = dp_i \wedge dx^i -\hat p_i dp^i \wedge dt + F - {\mathfrak F}, 
\ee
where $F = dA$ is the Maxwell tensor and ${\mathfrak F}= d{\mathfrak a}$ is the corresponding curvature for the Berry connection. Because  $dF = 0$ from Maxwell's equations, we have 
\bea 
d\omega_H &=& dF - d{\mathfrak F}\nonumber\\
& =&\phantom{  dF } - d{\mathfrak F}. 
\eea
Here, $d{\mathfrak F} \neq 0$ because of the  Berry monopole at the origin ${\bf p} = 0$. 

The Lie derivative of the Liouville volume form, using equation (\ref{EQ:form-liouville}), is simply 
\bea
{\mathcal L}_{\bf v} \Omega &=&  \frac{1}{3!} d\omega_H^3  = \frac{1}{2!} d\omega_H \wedge \omega_H^2 \nonumber \\
&=&  -\frac {1}{2} d{\mathfrak F} \wedge (dp_i \wedge dx^i -\hat p_i dp^i \wedge dt + F - {\mathfrak F})^2 \nonumber \\
&=&  -\frac {1}{2} d{\mathfrak F} \wedge F^2 = - \frac{1}{2} \left[ \epsilon^{ijk} \frac{\partial}{\partial p_i} \left( \frac{1}{2} {\mathfrak F}_{jk} \right) \right] \left[\frac{1}{2^2} \epsilon^{lmnq} F_{lm} F_{nq} \right] d^3p\, d^3x  \, dt \nonumber \\
&=& - \left(\frac{\partial {\mathfrak B}_i}{\partial p_i} \right) \frac{1}{2} \left[\epsilon^{0mnq} F_{0m} F_{nq} \right]  d^3p\, d^3x \, dt \nonumber \\
&=& -( \nabla_{\bf p}.{\mathfrak B} ) (-E_m) B_m \,  d^3p\,d^3x \, dt  \nonumber \\
&=&  ( \nabla_{\bf p}.{\mathfrak B} ) ({\mathbf E}.{\mathbf B})  d^3p\,d^3x  \, dt 
\eea

Comparison with equation  (\ref{EQ:t-dep-liouville})  gives
\be
\frac{\partial \sqrt{\omega}}{\partial t} + \frac{\partial \sqrt{\omega} \dot x_i}{\partial x_i} + \frac{\partial \sqrt{\omega} \dot p_i} {\partial p_i} = ( \nabla_{\bf p}.{\mathfrak B} ) ({\mathbf E}.{\mathbf B}) =2\pi  \delta^3(\mathbf p) ({\mathbf E}.{\mathbf B})
\ee
which is the result obtained in \cite{stephanov}. Defining the currents as in \cite{stephanov}, leads immediately  to
\be 
\partial_\mu J^\mu = \frac{1}{(2\pi)^2}  {\mathbf E}.{\mathbf B},
\ee
which is the expression for chiral anomaly in 3+1 dimensions.  Comparison with the labour  in \cite{stephanov,stone-dwivedi} shows the  compactness of  the present derivation.

  \section{Spin Chern number}
  \label{SEC:chern-character}

Here we compute the Chern number for the bundle of positive-energy eigenstates of  
 \be 
\hat H_{\bf p}= \sum_{i=1}^{2N+1} \Gamma_i p^i
\ee
over the $2N$-sphere in momentum space.

$\hat H$ has  eigenvalues $\pm |{\bf p}|$ and  
the projectors on to the positive and negative energy eigenspaces $V_+(\p)$, $V_-(\p)$  are 
\bea
P&=& \sum_\Lambda \ket{{\bf p},\alpha,+}\bra{{\bf p},\alpha,+}= \frac{1}{2}({\mathbb I}+ \hat p^i \Gamma_i), \nonumber\\
P^\perp&=& \sum_\Lambda \ket{{\bf p},\alpha,-}\bra{{\bf p},\alpha,-}= \frac{1}{2}({\mathbb I}- \hat p^i \Gamma_i),
\eea
respectively.
Here $\hat {\bf p}$ is the unit vector  ${\bf p}/|{\bf p}|$.

For any eigenvalue problem with positive and negative energy spaces the matrix-valued two-form $PdPdP$ 
has matrix elements only in $V_+(\p)$, and coincides there with the matrix elements of the Berry curvature $\mathfrak F=d\mathfrak a-i{\mathfrak a}^2$ divided by $i$. 
In other words 
\be 
iPdPdP= \sum_{\alpha,\beta} \ket{{\bf p},\alpha,+}{\mathfrak F}_{\alpha\beta} \bra{{\bf p},\beta,+}.
\ee
In our case 
\be
PdPdP
=\frac{i}{4}({\mathbb I}+ \hat p^i \Gamma_i) \frac 1{4i } [\Gamma_i,\Gamma_j] d\hat p^id\hat p^j.
\ee
For positive helicity we have  
\be
\Gamma_{1}\Gamma_{2}\cdots \Gamma_{{2N+1}} =  i^N {\mathbb I}_{2^N},
\ee
and can chose a  basis in which 
\be
\Gamma_{2N+1}= \left[\matrix{{\mathbb I}_{2^{N-1}} & 0\cr 0&-{\mathbb I}_{2^{N-1}}}\right].
\ee

To  relate   our set of $2N+1$ gamma matrices with the $2N$ needed for the representations of ${\rm Spin}(2N)$,  consider the fibre over a generic point in $\p$ space that we may as well take  ${\bf p}=(0,\ldots, 0,1)$.   Then, at that point
\be
dPdP= \sum_{i,j=1}^{2N} \frac{i}{2}\left[\matrix{\hat X_{ij,\sigma_+}& 0\cr0&  \hat X_{ij,\sigma_-}}\right]d\hat p^id\hat p^j, \quad 1\le i,j\le 2N,
\ee
where  
\be
\hat X_{ij,\sigma_+}= \frac{1}{4i} P[\Gamma_i,\Gamma_j]P,\quad \hat X_{ij,\sigma_-}= \frac{1}{4i} P^\perp[\Gamma_i,\Gamma_j]P^\perp,
\ee
 are the matrices representing the generator $X_{ij}$ in the two inequivalent spin representations of ${\rm Spin}(2N)$.  The projector $P\equiv P_{\sigma_+}=({\mathbb I}+ \Gamma_{2N+1})/2$  ensures that the    trace over the $2^N$ indices in $PdPdP$ coincides with the trace over the $2^{N-1}$ indices in the $\sigma_+$ representation.
and
\bea
&&\hspace{-2em}\tr_{\sigma_+}\left(\hat X_{i_1i_2,\sigma_+}\cdots \hat X_{i_{2N-1}i_{2N},\sigma_+}\right)\epsilon^{i_1i_2 \ldots i_{2N-1}i_{2N}}\nonumber\\
&=&(-i)^N 2^{-N} (2N)!\tr\{P_{\sigma_+} \Gamma_1\ldots \Gamma_{2N}\}\nonumber\\
&=&   2^{-N} (2N)! \frac 12  \tr\{{\mathbb I}_{2^N}\}\nonumber\\
&=&  (2N)!/2. 
\eea  
In the $\sigma_-$ representation, the RHS becomes $-(2N)!/2$. For any other point this  works the same way---except that $P_{\bf p}\equiv  \textstyle{\frac 12}({\mathbb I}+\hat p^i\Gamma_i)$  projects onto  an  equivalent set of ${\rm Spin}(2N)$ generators.

The Chern number is   given by the integral of the Chern-character class over the unit sphere in momentum space 
\bea
{\rm ch}_N(V_+)&\stackrel{\rm def}{=}&\frac{1}{N!} \left(\frac{1}{2\pi }\right)^{N}\int_{S^{2N}} \tr_{\sigma_+}({\mathfrak  F}^N)\nonumber\\
&=& \frac{1}{N!} \left(\frac{i}{2\pi }\right)^{N}
\int_{S^{2N}} \tr\{(PdPdP)^{N}\}
\eea

To evaluate this set 
\be
Z=\sum_{\mu=1}^{2N+1} \hat p^i \Gamma_i
\ee
so that  $Z=P-P^\perp$ and  
\bea
(2\times  4^N)\tr\{(iPdPdP)^{N}\}&=&i^N \tr\{Z(dZ)^{2N}\} \nonumber\\
&=& i^{2N} 2^N \epsilon_{i_1,\ldots,i_{2N+1}} \hat p^{i_1}d\hat p^{i_2}\cdots d\hat p^{i_{2N+1}}\nonumber\\
&=& (-1)^N 2^N (2N)! d[\hbox{Area on $S^{2N}$} ].
\eea
Thus 
\bea
\frac{i^N}{2^N (2N)!}\int_{S^{2N}} \tr\{ Z(dZ)^{2N}\}  &=& (-1)^N |S_{2N}|\nonumber\\
&=&   (-1)^N 2\pi^{N+1/2}/\Gamma\left(\frac{2N+1}{2}\right).
\eea
Here we have used a standard  formula for the surface area $|S^{2N}|$ of the $2N$-sphere. On further using $x\Gamma(x)=\Gamma(x+1)$ {\it etc.\/}\,  we find
\bea
{\rm ch}_N(V_+)&=&\frac{1}{N!} \left(\frac{1}{2\pi }\right)^{N}
\int_{S^{2N}} \tr\{(iPdPdP)^{N}\}\\
&=&
\frac {1}{2(N!)} \left(\frac{i}{8\pi}\right)^N \int_{S^{2n}} \tr\{Z(dZ)^{2N}\}
\nonumber\\
&=& (-1)^N.
\eea

\section{Classical and quantum traces for ${\rm Spin}(2N)$}
\label{SEC:trace-integrals}

The computation of the Chern character in appendix \ref{SEC:chern-character} required us to evaluate  the trace
\be
T_{\sigma_+}^{\rm quantum}=\epsilon^{i_1i_2\cdots i_{2N-1}  i_{2N}} \tr_{\sigma_+}\{\hat X_{i_1,i_2}\hat X_{i_3i_4}\cdots \hat X_{i_{2N-1}i_{2N}}\}
\ee
where
\be
\hat X_{ij}= \frac 1{4i}P_{\sigma_+}[\Gamma_i,\Gamma_j]P_{\sigma_+}
\ee
is the matrix representing the ${\rm Spin}(2N)$ generator $X_{ij}$ in the positive-helicity representation $\sigma_+$. 
Using 
\be
 \Gamma_{1}\Gamma_{2}\cdots \Gamma_{{2N+1}} =  i^N {\mathbb I}_{2^N}  
\ee
we found  that 
\be
T_{\sigma_+}^{\rm quantum}=(2N)!/2. 
\ee

For the classical calculation we need to evaluate  the corresponding phase space integral
\be
T^{\rm classical}_{\sigma_+}=\int_{{\mathcal O}_{\sigma_+}} \epsilon^{i_1i_2\cdots i_{2N-1}  i_{2N}}  {\mathfrak S}_{i_1i_2}  {\mathfrak S}_{i_3i_4} \cdots {\mathfrak S}_{ i_{2N-1}i_{2N}}\d\bar \mu_{\sigma_+}
\ee
To do this we 
begin by summarizing some basic facts about the Lie algebra of ${\rm Spin}(2N)$, which is the double cover of  ${\rm SO}(2N)$ and so shares the same Lie algebra. 

The Hermitian matrix generators in the defining vector representation of ${\rm SO}(2N)$   are $X_{ij}= -i(e_{ij}-e_{ji})$. Here  $e_{ij}$  is the matrix with unity at site $i,j$,  so that 
\be
e_{ij}e_{kj}= \delta_{jk}e_{il}.
\ee
Thus the commutation relations are 
\be
[X_{ij},X_{mn}]=i(\delta_{mi}X_{jn}- \delta_{mj}X_{in} -X_{mi}\delta_{jn}+ X_{mj}\delta_{in}).
\ee
These relations  can be understood as saying  that $X_{mn}$ transforms a skew 2-tensor under the adjoint action of the group.

With traces taken in the defining vector representation, we have
\be
\tr\{X_{ij}X_{kl}\}=2(\delta_{ik}\delta_{jl}-\delta_{il}\delta_{jk}).
\ee
If we take as our basis set the $X_{ij}$ with $i<j$, the second term in this trace is always zero. The  ${\rm Spin}(2N)$ analogue of the metric  (\ref{EQ:lie-metric})  is therefore 
\be
g_{ij,kl} = \tr\{X_{ij}X_{kl}\}=2\delta_{ik}\delta_{jl}.  
\ee
The Cartan algebra is generated by
\be
h_n=X_{2n-1,2n},\quad n=1,\ldots, N,
\ee
and a general weight will be of the form ${\mathfrak s}=(m_1,m_2,\ldots,m_N)$ where $m_n$ is the eigenvalue of $h_n$.
The  roots are $\pm {\bf e}_i \pm {\bf e}_j$ where ${\bf e}_1=(1,0,\ldots, 0)$ {\it etc\/}.
In this  basis the  $N$  fundamental weights are 
\bea
{\bm \omega}_1&=&\phantom{\textstyle{\frac 12 }}(1,0,\ldots,0,0,0),\nonumber\\
{\bm \omega}_2&=&\phantom{\textstyle{\frac 12 }}(1,1, \ldots,0,0,0),\nonumber\\
&& \quad \qquad \vdots\nonumber\\
{\bm \omega}_{N-2}&=&\phantom{\textstyle{\frac 12 }}(1,1,\ldots 1,0,0),\nonumber\\
{\bm \omega}_{N-1}\equiv {\bm \sigma_+}&=&\textstyle{\frac 12 }(1,1,\ldots,1,1,1), \nonumber\\
{\bm \omega}_N\equiv  {\bm \sigma_-}&=&\textstyle{\frac 12}(1,1,\ldots1, 1,-1).\nonumber
\eea
The highest weight of any unitary representation is a positive-integer linear combination of the fundamental weights.  The weight ${\bm \omega}_1$ is  that of  the $2N$-dimensional defining vector representation of ${\rm SO}(2N)$.
The last two fundamental weights are those of the  two  inequivalent spin representations of ${\rm Spin}(2N)$.  They are not representations of ${\rm SO}(2N)$. 
The Weyl vector  ${\bm \rho}$ is defined to be half the sum of the positive roots, or equivalently the sum of the fundamental weights, and is therefore given by
\be
{\bm \rho}=(N-1,N-2,\ldots,1,0).
\ee


Consider the representation with highest weight $\bf {\mathfrak s}$ and highest weight vector $\ket{\mathfrak s}$.
The Cartan algebra element  $\beta_{\mathfrak s}$ of equation (\ref{EQ:beta-s})  is defined so that 
\be
\tr \{\beta_{\mathfrak s} X\} = \eval{\mathfrak s}{X}{\mathfrak s}
\ee
for any $X$ in the Lie algebra.
In the representation $\sigma_+$ with weight ${\mathfrak s}={\bm \sigma_+}$ we have
\be
\eval{\sigma_+}{X_{12}}{\sigma_+}=\eval{\sigma_+}{X_{34}}{\sigma_+}=\ldots=\eval{\sigma_+}{X_{2N-1,2N}}{\sigma_+}= \textstyle{\frac 12}
\ee
with all non-Cartan elements giving zero. 
Thus 
\be
\beta_{\sigma_+}= \frac 14(X_{12}+X_{34}+\cdots+X_{2N-1,2N}).
\ee
The  orbit coordinates are  ${\mathfrak S}^{ij}$, $i<j$ given by
\be
\sum_{i<j} {\mathfrak S}^{ij}X_{ij} = \sigma \beta_{\sigma_+} \sigma^{-1}.
\ee
In the Weyl chamber ${\mathfrak S}^{12}={\mathfrak S}^{34}=\cdots = {\mathfrak S}^{2N-1,2N}=1/4$ with all other components being zero. Using the metric to lower the indices gives 
\be
{\mathfrak S}_{12}={\mathfrak S}_{34}=\cdots = {\mathfrak S}_{2N-1,2N}=1/2.
\ee
We know therefore that 
\be
C_{\sigma_+}^{\rm geom}\stackrel{\rm def}{=}\epsilon^{i_1\ldots i_{2N}} {\mathfrak S}_{i_1i_2}\cdots {\mathfrak S}_{i_{2N-1}i_{2N}}=N!
\ee
at the point in which ${\mathcal O}_{\sigma_+}$ intersects the Weyl chamber. Now $\epsilon^{i_1\ldots i_{2N}} $ is an invariant tensor for the adjoint action and so  $C_{\sigma_+}^{\rm geom}$ is a classical Casimir invariant and  takes the same  value everywhere on the orbit.
Thus we find that 
\be
T_{\sigma_+}^{\rm classical}\stackrel{\rm ?}{=}\int_{{\mathcal O}_{\sigma_+}}\epsilon^{i_1\ldots i_{2N}} {\mathfrak S}_{i_1i_2}\cdots {\mathfrak S}_{i_{2N-1}i_{2N}}d \bar\mu_{\sigma_+}= {\rm Vol}({\mathcal O}_{\sigma_+}) \,N!.
\label{EQ:badtrace}
\ee
In the case of a representation with weight ${\mathfrak s}=(m_1,m_2,\ldots, m_n)$  we have ${\mathfrak S}_{12}=m_1,$ ${\mathfrak S}_{34}=m_2$, {\it etc\/}.\ so this   integral generalizes to 
\be
T^{\rm classical}_{\mathfrak s}\stackrel{\rm ?}{=}\int_{{\mathcal O}_{\mathfrak s}}\epsilon^{i_1\ldots i_{2N}} {\mathfrak S}_{i_1i_2}\cdots {\mathfrak S}_{i_{2N-1}i_{2N}} d\bar\mu_{\mathfrak s}= {\rm Vol}({\mathcal O}_{\mathfrak s})  2^{N} N! \, m_1m_2\cdots m_N.
\label{EQ:secondbadtrace}
\ee
The reason for the ``?'' over the equal signs is that  ${\rm Vol}({\mathcal O}_{\mathfrak s}) $ should correspond to $\tr_{\mathfrak S}({\mathbb I}_{{\rm dim}({\mathfrak s})})= {\rm dim}({\mathfrak s})$, and since ${\rm dim}(\sigma_+)=2^{N-1}$ the right-hand side of (\ref{EQ:badtrace}) 
would give $2^{N-1} N!$.  This  is not a very good approximation $(2N)!/2$.

It turns out that the na{\"i}ve orbit volumes  are often poor  approximations to the dimension of the representation---particularly on the edges of the Weyl Chamber where the orbit degenerates and reduces in dimension.  
It is, however,  a corollary of the  Kirillov character formula \cite{kirillov} that  the volume of the orbit corresponding the Weyl-shifted weight   ${\mathfrak s}\to {\mathfrak s}+\bm \rho$ gives the correct dimension:
\be
\int_{{\mathcal O}_{{\mathfrak s}+\rho}}d \bar\mu_{{\mathfrak s}+\rho}= {\rm Vol}({\mathcal O}_{{\mathfrak s}+\rho})= {\rm dim}({\mathfrak s}).
\ee
The Weyl shift is a quantum correction that accounts for normal-ordering of generators in the Casimir operators. In the spin-$j$ representation of ${\rm SU}(2)$ for example, the Weyl shift changes the orbit volume from $2j$ to $2j+1$.  For large spins this is a negligible effect, but it is  important for the fundamental spin representation in high dimensions. 

Suppose  we make a similar Weyl shift in  (\ref{EQ:secondbadtrace}) so that it is replaced by  
\be
T_{\mathfrak s}^{\rm classical} =\int_{{\mathcal O}_{\mathfrak s +\rho}}\epsilon^{i_1\ldots i_{2N}} {\mathfrak S}_{i_1i_2}\cdots {\mathfrak S}_{i_{2N-1}i_{2N}} d\bar \mu_{\mathfrak s +\rho}
={\rm dim}({\mathfrak s}) 2^{N}  N!  (m_1+N-1)(m_2+N-2)\cdots m_1.
\label{EQ:goodtrace}
\ee
The formul{\ae}\  for the ${\rm Spin}(2N)$  Casimir operators  in \cite{perelemov} now show that the classical trace formula (\ref{EQ:goodtrace}) coincides with the quantum trace for all $\mathfrak s$.
In particular   (\ref{EQ:badtrace}) becomes   
\bea
T_{\sigma_+}^{\rm classical}=\int_{{\mathcal O}_{{\mathfrak s}+\rho}}\epsilon^{i_1\ldots i_{2N}} {\mathfrak S}_{i_1i_2}\cdots {\mathfrak S}_{i_{2N-1}i_{2N}}d \bar\mu_{\sigma_++\rho}&=& 2^{N-1} (2^N N!)
(N-\textstyle{\frac 12})(N-\textstyle{\frac 32 })\cdots \textstyle{\frac 12}\nonumber\\
&=& (2N)!/2,\nonumber\\
&=& T_{\sigma_+}^{\rm quantum}.
\label{EQ:badtrace}
\eea

 There are other terms involving ${\mathfrak F}^n$, $n<N$, in the expansion of $\omega_H$, but the structure of ${\mathfrak F}$ (skew-symmetric form indices coincide  with the Lie algebra  indices) requires all pairs of $i,j$ indices on the ${\mathfrak S}_{ij}$ in these products to be distinct.
Now  $\epsilon^{i_1\ldots i_{2N}} $ is the only adjoint-action invariant tensor that allows for distinct indices. All other invariant tensors are sums of products of $\delta_{ij}$, and so give zero unless pairs of  indices coincide.  As a consequence no other trace, either quantum or classical, can make a  contribution to the anomaly.

\end{document}